\documentclass[12pt]{article}
\usepackage{amsmath,amsthm}
\usepackage{eucal}
\usepackage{amsfonts}
\newcommand{\ar}{\renewcommand{\arraystretch}{1}} 
\gdef\C{\Bbb C}

\gdef\dS{\Bbb S}

\gdef\F{\Bbb F}

\DeclareMathOperator{\fD}{\mathfrak{D}}

\DeclareMathOperator{\bA}{{\bf A}}

\DeclareMathOperator{\Sym}{Sym}

\newcommand{\cA}{\mathcal{A}}

\newcommand{\sI}{{\sf I}}

\newcommand{\bx}{{\bf x}}

\newcommand{\bF}{{\bf F}}
\newcommand{\bE}{{\bf E}}
\newcommand{\bH}{{\bf H}}

\newcommand{\fC}{\mathfrak{C}}
\newcommand{\fR}{\mathfrak{R}}
\newcommand{\fH}{\mathfrak{H}}

\newcommand{\fG}{\mathfrak{G}}

\newcommand{\cl}{C\kern -0.2em \ell}

\newcommand{\e}{\mbox{\bf e}}

\begin{document}
\title{About Algebraic Foundations of Majorana-Oppenheimer Quantum
Electrodynamics and de Broglie-Jordan Neutrino Theory of Light}
\author{V.~V. Varlamov\thanks{Department of Mathematics, Siberia State
University of Industry, Kirova 42, Novokuznetsk 654007, Russia.}}
\date{}
\maketitle
\begin{abstract}
An algebraic description of basic physical fields (neutrino field,
electron--positron field and electromagnetic field) is studied. It is
shown that the electromagnetic field can be described within a quotient
representation of the proper orthochronous Lorentz group. The relation of
such a description with Majorana--Oppenheimer quantum electrodynamics and
de Broglie--Jordan neutrino theory of light is discussed.
\end{abstract}
\bigskip
Many years ago Bogoliubov and Shirkov \cite{BS93} pointed out that among
all physical fields the electromagnetic field (beyond all shadow of doubt
the main physical field) is quantized with the most difficulty. In the
standard Gupta--Bleuler approach an unobservable magnitude 
(electromagnetic four--potential $\bA$)
is quantized. At this point, the four--potential has four degrees of
freedom, but in nature there are only two degrees of freedom for a
photon field (left and right handed polarizations). Besides, the electromagnetic
four--potential is transformed within $(1/2,1/2)$--representaion of the
homogeneous Lorentz group and, therefore, in accordance with 
a well-known Weinberg Theorem \cite{Wein} the field described by $\bA$ has 
a null helicity, that also contradicts with experience. Moreover,
at the present time electromagnetic field is understood as a
`gauge field' that gives rise to a peculiar opposition with other physical
fields called by this reason as `matter fields'.

With the aim of overcoming this unnatural opposition all the physical
fields should be considered on an equal footing. In this paper we present
an algebraic construction of the most fundamental physical fields such as
neutrino field, electron--positron field and electromagnetic field.
Our consideration based mainly on the relation between Clifford algebras and
Lorentz group (all mathematical background contained in 
\cite{Var99,Var00,Var03,Var01a,Var01}).
In \cite{Var01} all the Clifford algebras are understood as
`algebraic coverings' of finite--dimensional representations of the proper
Lorentz group $\fG_+$. In \cite{Var01} it has been shown that there is a
following classification:\\[0.2cm]
{\bf I}. Complex representations.
\begin{description}
\item[1)] Representations $\fC^{l_0+l_1-1,0}\leftrightarrow\C_n$ with
the field $(j,0)$, where $j=\frac{l_0+l_1-1}{2}$.
\item[2)] Representations 
$\fC^{0,l^\prime_0-l^\prime_1+1}\leftrightarrow\overset{\ast}{\C}_n$ 
with the field
$(0,j^\prime)$, where $j^\prime=\frac{l^\prime_0-l^\prime_1+1}{2}$.
\item[3)] Representations 
$\fC^{l_0+l_1-1,l^\prime_0-l^\prime_1+1}\leftrightarrow
\C_n\otimes\overset{\ast}{\C}_n$
with the field $(j,j^\prime)$.
\item[4)] Representations
$\fC^{l_0+l_1-1,0}\oplus\fC^{0,l_0-l_1+1}\leftrightarrow
\C_n\oplus\overset{\ast}{\C}_n$ with the field $(j,0)\oplus(0,j)$, 
$j=j^\prime$.
\item[5)] Quotient representations
${}^\chi\fC^{l_0+l_1-1,0}\cup{}^\chi\fC^{0,l_0-l_1+1}\leftrightarrow
{}^\epsilon\C_n\cup{}^\epsilon\overset{\ast}{\C}_n$ with the field
$(j,0)\cup(0,j)$.
\end{description}
{\bf II}. Real representations.
\begin{description}
\item[6)] Real representations $\fR^{l_0}_{0,2}\leftrightarrow\cl_{p,q}$,
$p-q\equiv 0,2\pmod{8}$, with the field $[j]$, where $j=\frac{l_0}{2}$,
$l_0=\frac{p+q}{4}$.
\item[7)] Quaternionic representations
$\fH^{l_0}_{4,6}\leftrightarrow\cl_{p,q}$, $p-q\equiv 4,6\pmod{8}$,
with the field $[j]$.
\item[8)] Quotient representations
${}^\chi\fD^{l_0}\cup{}^\chi\fD^{l_0}\leftrightarrow
{}^\epsilon\cl_{p,q}\cup{}^\epsilon\cl_{p,q}$ with the field
$[j]\cup[j]$, where ${}^\chi\fD^{l_0}=\{{}^\chi\fR^{l_0}_{0,2},
{}^\chi\fH^{l_0}_{4,6}\}$.
\end{description} 
Here the numbers $l_0$ and $l_1$ define the finite--dimensional
representation in the Gel'fand--Naimark representation theory of the Lorentz
group \cite{GMS,Nai58}. In its turn, quotient representations correspond to
the type $n\equiv 1\pmod{2}$ of $\C_n$ (or to the types
$p-q\equiv 1,5\pmod{8}$ for the real representations). Over the field
$\F=\C$ these representations obtained in the result of the following
decomposition
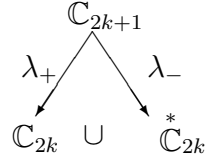
\[
\unitlength=0.5mm
\begin{picture}(70,50)
\put(35,40){\vector(2,-3){15}}
\put(35,40){\vector(-2,-3){15}}
\put(28.25,42){$\C_{2k+1}$}
\put(16,28){$\lambda_{+}$}
\put(49.5,28){$\lambda_{-}$}
\put(13.5,9.20){$\C_{2k}$}
\put(52.75,9){$\stackrel{\ast}{\C}_{2k}$}
\put(32.5,10){$\cup$}
\end{picture}
\]
Here central idempotents
\[
\lambda^+=\frac{1+\varepsilon\e_1\e_2\cdots\e_{2k+1}}{2},\quad
\lambda^-=\frac{1-\varepsilon\e_1\e_2\cdots\e_{2k+1}}{2},
\]
where
\[
\varepsilon=\begin{cases}
1,& \text{if $k\equiv 0\pmod{2}$},\\
i,& \text{if $k\equiv 1\pmod{2}$}
\end{cases}
\]
satisfy the relations $(\lambda^+)^2=\lambda^+$, $(\lambda^-)^2=\lambda^-$,
$\lambda^+\lambda^-=0$. Thus, we have a decomposition of the initial
algebra $\C_{2k+1}$ into a direct sum of two mutually annihilating simple
ideals: $\C_{2k+1}\simeq\frac{1}{2}(1+\varepsilon\omega)\C_{2k+1}\oplus
\frac{1}{2}(1-\varepsilon\omega)\C_{2k+1}$. Each of the ideals
$\lambda^{\pm}\C_{2k+1}$ is isomorphic to the subalgebra 
$\C_{2k}\subset\C_{2k+1}$. In accordance with Chisholm and Farwell \cite{CF97}
the idempotents $\lambda^+$ and $\lambda^-$ can be identified with
helicity projection operators which distinguish left and right handed
spinors. The Chisholm--Farwell notation for $\lambda^\pm$ we will widely
use below.

The first simplest case of such a decomposition is presented by the algebra
$\C_3$ related with the neutrino field. Indeed,
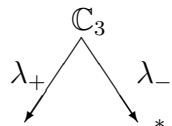
\[
\unitlength=0.5mm
\begin{picture}(70,50)
\put(35,40){\vector(2,-3){15}}
\put(35,40){\vector(-2,-3){15}}
\put(32.25,42){$\C_{3}$}
\put(16,28){$\lambda_{+}$}
\put(49.5,28){$\lambda_{-}$}
\put(13.5,9.20){$\C_{2}$}
\put(52.75,9){$\stackrel{\ast}{\C}_{2}$}
\put(32.5,10){$\cup$}
\end{picture}
\]
here central idempotents
\begin{equation}\label{Cent}
\lambda_{-}=\frac{1-i\e_1\e_2\e_3}{2},\quad
\lambda_{+}=\frac{1+i\e_1\e_2\e_3}{2} 
\end{equation}
in accordance with \cite{CF97}
can be identified with helicity projection operators. In such a way, we have
two helicity states describing by the quotient algebras 
${}^\epsilon\C_2$ and ${}^\epsilon\overset{\ast}{\C}_2$, and a full
neutrino--antineutrino algebra is ${}^\epsilon\C_2\cup{}^\epsilon
\overset{\ast}{\C}_2$ (cf. electron--positron algebra
$\C_2\oplus\overset{\ast}{\C}_2$). 

Let $\varphi\in\C_3$ be an algebraic spinor of the form
(sometimes called operator spinor, see \cite{FRO90a})
\begin{equation}\label{Neut1}
\varphi=a^0+a^1\e_1+a^2\e_2+a^3\e_3+a^{12}\e_1\e_2+a^{13}\e_1\e_3+
a^{23}\e_2\e_3+a^{123}\e_1\e_2\e_3.
\end{equation}
Then it is easy to verify that spinors
\begin{equation}\label{Neut2}
\varphi^+=\lambda_+\varphi=\frac{1}{2}(1+i\e_1\e_2\e_3)\varphi,\quad
\varphi^-=\lambda_-\varphi=\frac{1}{2}(1-i\e_1\e_2\e_3)\varphi
\end{equation}
are mutually orthogonal, $\varphi^+\varphi^-=0$, since 
$\lambda_+\lambda_-=0$, and also $\varphi^+\in\C_2$,
$\varphi^-\in\overset{\ast}{\C}_2$. Further, it is obvious that a spinspace
of the algebra ${}^\epsilon\C_2\cup{}^\epsilon\overset{\ast}{\C}_2$ is
$\dS_2\cup\dot{\dS}_2$. It should be noted here that structures of the
spinspaces $\dS_2\cup\dot{\dS}_2$ and $\dS_2\oplus\dot{\dS}_2$
are different. Indeed,
\[
\dS_2\cup\dot{\dS}_2=\ar\begin{pmatrix}
\left[00,\dot{0}\dot{0}\right] & \left[01,\dot{0}\dot{1}\right]\\
\left[10,\dot{1}\dot{0}\right] & \left[11,\dot{1}\dot{1}\right]
\end{pmatrix},\quad
\dS_2\oplus\dot{\dS}_2=\begin{pmatrix}
00 & 01 & & \\
10 & 11 & & \\
   &    &\dot{0}\dot{0} & \dot{0}\dot{1}\\
   &    &\dot{1}\dot{0} & \dot{1}\dot{1}
\end{pmatrix}.
\]
Under action of the pseudoautomorphism $\cA\rightarrow\overline{\cA}$
(charge conjugation $C$, see \cite{Var01}) spinspace $\dS_2\cup\dot{\dS}_2$
take a form
\[
\dot{\dS}_2\cup\dS_2=\ar\begin{pmatrix}
\left[\dot{0}\dot{0},00\right] & \left[\dot{0}\dot{1},01\right]\\
\left[\dot{1}\dot{0},10\right] & \left[\dot{1}\dot{1},11\right]
\end{pmatrix}.
\]
Since spinor representations of the quotient algebras ${}^\epsilon\C_2$ and
${}^\epsilon\overset{\ast}{\C}_2$ are defined in terms of Pauli matrices
$\sigma_i$, then the algebraic spinors $\varphi^+\in{}^\epsilon\C_2$ and
$\varphi^-\in{}^\epsilon\overset{\ast}{\C}_2$ correspond to spinors
$\xi^{\alpha_i}\in\dS_2$ and $\xi^{\dot{\alpha}_i}\in\dot{\dS}_2$
($i=0,1$). Hence we have Weyl equations
\begin{equation}\label{Weyl}
\left(\frac{\partial}{\partial x^0}-\boldsymbol{\sigma}
\frac{\partial}{\partial\bx}\right)\xi^{\alpha}=0,\quad
\left(\frac{\partial}{\partial x^0}+\boldsymbol{\sigma}
\frac{\partial}{\partial\bx}\right)\xi^{\dot{\alpha}}=0.
\end{equation}
Therefore, two--component Weyl theory can be naturally formulated within
quotient representation ${}^\chi\fC^{1,0}_c\cup{}^\chi\fC^{0,-1}_c$ of the
group $\fG_+$. Further, in virtue of an isomorphism 
$\C_2\simeq\cl_{3,0}\simeq\cl^+_{1,3}$ ($\cl_{1,3}$ is the space--time
algebra) the spinor field of the quotient representation ${}^\chi\fC^{0,-1}_c$
(${}^\chi\fC^{1,0}_c$) can be expressed via the Dirac--Hestenes spinor
field $\phi(x)\in\cl_{3,0}$ \cite{Hest66,Hest67,Lou93}. 
Indeed, the Dirac--Hestenes spinor is represented by a following
biquaternion number
\begin{equation}
\phi=a^0+a^{01}\gamma_0\gamma_1+a^{02}\gamma_0\gamma_2+a^{03}\gamma_0\gamma_3+
a^{12}\gamma_1\gamma_2+a^{13}\gamma_1\gamma_3+a^{23}\gamma_2\gamma_3
+a^{0123}\gamma_0\gamma_1\gamma_2\gamma_3,
\label{173}
\end{equation}
or using $\gamma$--matrix basis
\begin{equation}\label{173'}
\ar\gamma_0=\begin{pmatrix}
I & 0\\
0 & -I
\end{pmatrix},\;\;\Gamma_1=\begin{pmatrix}
0 & \sigma_1\\
-\sigma_1 & 0
\end{pmatrix},\;\;\Gamma_2=\begin{pmatrix}
0 & \sigma_2\\
-\sigma_2 & 0
\end{pmatrix},\;\;\Gamma_3=\begin{pmatrix}
0 & \sigma_3\\
-\sigma_3 & 0
\end{pmatrix},
\end{equation}
we can write (\ref{173}) in the matrix form
\begin{equation}\label{174}
\ar\phi=\begin{pmatrix}
\phi_1 & -\phi^\ast_2 & \phi_3 & \phi^\ast_4 \\
\phi_2 & \phi^\ast_1 & \phi_4 & -\phi^\ast_3\\
\phi_3 & \phi^\ast_4 & \phi_1 & -\phi^\ast_2\\
\phi_4 & -\phi^\ast_3 & \phi_2 & \phi^\ast_1
\end{pmatrix},
\end{equation}
where
\[
\phi_1=a^0-ia^{12},\quad
\phi_2=a^{13}-ia^{23},\quad
\phi_3=a^{03}-ia^{0123},\quad
\phi_4=a^{01}+ia^{02}.
\]
From (\ref{Neut1})--(\ref{Neut2}) and (\ref{173}) it is easy to see that
spinors $\varphi^+$ and $\varphi^-$ are algebraically equivalent to the
spinor $\phi\in\C_2\simeq\cl_{3,0}$. Further, since $\phi\in\cl^+_{1,3}$,
then actions of the antiautomorphisms $\cA\rightarrow\widetilde{\cA}$ and
$\cA\rightarrow\widetilde{\cA^\star}$ on the field $\phi$ are equivalent.
On the other hand, in accordance with Feynman--Stueckelberg interpretation,
time reversal for the chiral field is equivalent to charge conjugation
(particles reversed in time are antiparticles). Thus, for the field
$\phi\in{}^\chi\fC^{0,-1}_c$ we have $C\sim T$ and, therefore, this field
is $CP$--invariant.

The spinor (\ref{173}) (or (\ref{174})) satisfies the Dirac--Hestenes
equation
\begin{equation}\label{DH}
\partial\phi\gamma_2\gamma_1-\frac{mc}{\hbar}\phi\gamma_0=0,
\end{equation}
where $\partial=\gamma^\mu\frac{\partial}{\partial x^\mu}$ is the Dirac
operator. Let us show that a massless Dirac--Hestenes equation
\begin{equation}\label{DHM}
\partial\phi\gamma_2\gamma_1=0
\end{equation}
describes the neutrino field. Indeed, the matrix 
$\gamma_0\gamma_1\gamma_2\gamma_3$ commutes with all the elements
of the biquaternion (\ref{173}) and, therefore, 
$\gamma_0\gamma_1\gamma_2\gamma_3$ is equivalent
to the volume element $\omega=\e_1\e_2\e_3$ of the biquaternion algebra
$\cl_{3,0}$. In such a way, we see that idempotents
\[
P_+=\frac{1+\gamma_5}{2},\quad P_-=\frac{1-\gamma_5}{2}
\]
cover the central idempotents (\ref{Cent}), where 
$\gamma_5=-i\gamma_0\gamma_1\gamma_2\gamma_3$.
Further, from (\ref{DHM}) we
obtain
\[
P_\pm\gamma^\mu\frac{\partial}{\partial x^\mu}\phi\gamma_2\gamma_1=
\gamma^\mu P_mp\frac{\partial}{\partial x^\mu}\phi\gamma_2\gamma_1=0,
\]
that is, there are two separated equations for 
$\phi^\pm=P_\pm\phi\gamma_2\gamma_1$:
\begin{equation}\label{DHM2}
\gamma^\mu\frac{\partial}{\partial x^\mu}\phi^\pm=0,
\end{equation}
where
\[
\phi^\pm=\frac{1}{2}(1\pm\gamma_5)\phi\gamma_2\gamma_1=\frac{i}{2}
\ar\begin{pmatrix}
\phi_1\mp\phi_3 & \phi^\ast_2\pm\phi^\ast_4 & \phi_3\mp\phi_1 &
-\phi^\ast_4\mp\phi^\ast_2\\
\phi_2\mp\phi_4 & -\phi^\ast_1\mp\phi^\ast_3 & \phi_4\mp\phi_2 &
\phi^\ast_3\pm\phi^\ast_1\\
\mp\phi_1+\phi_2 & \mp\phi^\ast_2-\phi^\ast_4 & \mp\phi_3+\phi_1 &
\pm\phi^\ast_4+\phi^\ast_2\\
\mp\phi_2+\phi_4 & \pm\phi^\ast_1+\phi^\ast_3 & \mp\phi_4+\phi_2 &
\mp\phi^\ast_3-\phi^\ast_1
\end{pmatrix}
\]
Therefore, each of the functions $\phi^+$ and $\phi^-$ contains only four
independent components and in the split form we have
\[
\phi^+=\ar\begin{pmatrix}
\psi_1 & \psi_2 & \psi_3 & \psi_4\\
-\psi_1 & -\psi_2 & -\psi_3 & -\psi_4
\end{pmatrix},\quad
\phi^-=\begin{pmatrix}
\psi_5 & \psi_6 & \psi_7 & \psi_8\\
\psi_5 & \psi_6 & \psi_7 & \psi_8
\end{pmatrix},
\]
where
\begin{gather}\ar
\psi_1=\frac{i}{2}\begin{pmatrix}
\phi_1-\phi_3\\
\phi_2-\phi_4
\end{pmatrix},\;\;
\psi_2=\frac{i}{2}\begin{pmatrix}
\phi^\ast_2+\phi^\ast_4\\
-\phi^\ast_1-\phi^\ast_3
\end{pmatrix},\;\;
\psi_3=\frac{i}{2}\begin{pmatrix}
\phi_3-\phi_1\\
\phi_4-\phi_2
\end{pmatrix},\;\;
\psi_4=\frac{i}{2}\begin{pmatrix}
-\phi^\ast_4-\phi^\ast_2\\
\phi^\ast_3+\phi^\ast_1
\end{pmatrix},\nonumber\\
\psi_5=\frac{i}{2}\ar\begin{pmatrix}
\phi_1+\phi_3\\
\phi_2+\phi_4
\end{pmatrix},\;\;
\psi_6=\frac{i}{2}\begin{pmatrix}
\phi^\ast_2-\phi^\ast_4\\
-\phi^\ast_1+\phi^\ast_3
\end{pmatrix},\;\;
\psi_7=\frac{i}{2}\begin{pmatrix}
\phi_3+\phi_1\\
\phi_4+\phi_2
\end{pmatrix},\;\;
\psi_8=\frac{i}{2}\begin{pmatrix}
-\phi^\ast_4+\phi^\ast_2\\
\phi^\ast_3-\phi^\ast_1
\end{pmatrix}.\nonumber
\end{gather}
Thus, in the $\gamma$--matrix basis we obtain from (\ref{DHM2})
\[
\left(\frac{\partial}{\partial x^0}-\boldsymbol{\sigma}
\frac{\partial}{\partial\bx}\right)\psi_i=0,\quad
\left(\frac{\partial}{\partial x^0}+\boldsymbol{\sigma}
\frac{\partial}{\partial\bx}\right)\psi_{i+4}=0,\quad(i=1,2,3,4)
\]
These equations are equivalent to Weyl equations (\ref{Weyl})
for neutrino field and, therefore, by analogy 
with the Dirac--Hestenes equations for $m\neq 0$
should be called {\it Weyl--Hestenes equations for neutrino field}.

The Dirac electron--positron field
$(1/2,0)\oplus(0,1/2)$ corresponds to the algebra $\C_2\oplus
\overset{\ast}{\C}_2$. It should be noted that the Dirac algebra
$\C_4$ considered as a tensor product $\C_2\otimes\C_2$ 
(or $\C_2\otimes\overset{\ast}{\C}_2$) 
gives rise to spintensors $\xi^{\alpha_1\alpha_2}$
(or $\xi^{\alpha_1\dot{\alpha}_1}$), but it contradicts with the usual
definition of the Dirac bispinor as a pair 
$(\xi^{\alpha_1},\xi^{\dot{\alpha}_1})$. Therefore, the Clifford algebra
associated with the Dirac field is $\C_2\oplus\overset{\ast}{\C}_2$, and
a spinspace of this sum in virtue of unique decomposition
$\dS_2\oplus\dot{\dS}_2=\dS_4$ ($\dS_4$ is a spinspace of $\C_4$) allows to
define $\gamma$--matrices in the Weyl basis.

In common with other massless fields (such as the neutrino field
$(1/2,0)\cup(0,1/2)$) the Maxwell electromagnetic field is also described
within the quotient representations of the Lorentz group. In accordance
with Theorem 4 in \cite{Var01} the photon field can be described by a
quotient representation of the class 
${}^\chi\fC^{2,0}_{a_1}\cup{}^\chi\fC^{0,-2}_{a_1}$. This representation
admits time reversal $T$ and an identical charge conjugation $C\sim\sI$
that corresponds to truly neutral particles (see Theorem 3 in \cite{Var01}).
The quotient algebra ${}^\epsilon\C_4\cup{}^\epsilon\overset{\ast}{\C}_4$
associated with the Maxwell field $(1,0)\cup(0,1)$ is obtained in the 
result of an homomorphic mapping $\epsilon:\,C_5\rightarrow\C_4$. Indeed,
for the algebra $\C_5$ we have a decomposition
\[
\unitlength=0.5mm
\begin{picture}(70,50)
\put(35,40){\vector(2,-3){15}}
\put(35,40){\vector(-2,-3){15}}
\put(32.25,42){$\C_{5}$}
\put(16,28){$\lambda_{+}$}
\put(49.5,28){$\lambda_{-}$}
\put(13.5,9.20){$\C_{4}$}
\put(52.75,9){$\stackrel{\ast}{\C}_{4}$}
\put(32.5,10){$\cup$}
\end{picture}
\]
where the central idempotents
\[
\lambda_+=\frac{1+\e_1\e_2\e_3\e_4\e_5}{2},\quad
\lambda_-=\frac{1-\e_1\e_2\e_3\e_4\e_5}{2}
\]
correspond to the helicity projection operators of the Maxwell field.
As known, for the photon there are two helicity states: left and right
handed polarizations.

Let $\varphi\in\C_5$ be an algebraic spinor of the form
\[
\varphi=a^0+\sum^5_{i=1}a^i\e_i+\sum^5_{i,j=1}a^{ij}\e_i\e_j+
\sum^5_{i,j,k=1}a^{ijk}\e_i\e_j\e_k+
\sum^5_{i,j,k,l=1}a^{ijkl}\e_i\e_j\e_k\e_l+a^{12345}\e_1\e_2\e_3\e_4\e_5,
\]
then the spinors 
\[
\varphi^+=\lambda_+\varphi=\frac{1}{2}(1+\e_1\e_2\e_3\e_4\e_5)\varphi,\quad
\varphi^-=\lambda_-\varphi=\frac{1}{2}(1-\e_1\e_2\e_3\e_4\e_5)\varphi
\]
are mutually orthogonal, $\varphi^+\varphi^-=0$, and $\varphi^+\in\C_4$,
$\varphi^-\in\overset{\ast}{\C}_4$. The spinspace of the algebra
${}^\epsilon\C_4\cup{}^\epsilon\overset{\ast}{\C}_4$ has a form
\[
{\renewcommand{\arraystretch}{1.4}
\dS_4\cup\dot{\dS}_4=\begin{pmatrix}
\left[0000,\dot{0}\dot{0}\dot{0}\dot{0}\right] &
\left[0001,\dot{0}\dot{0}\dot{0}\dot{1}\right] &
\left[0010,\dot{0}\dot{0}\dot{1}\dot{0}\right] &
\left[0011,\dot{0}\dot{0}\dot{1}\dot{1}\right] \\
\left[0100,\dot{0}\dot{1}\dot{0}\dot{0}\right] &
\left[0101,\dot{0}\dot{1}\dot{0}\dot{1}\right] &
\left[0110,\dot{0}\dot{1}\dot{1}\dot{0}\right] &
\left[0111,\dot{0}\dot{1}\dot{1}\dot{1}\right] \\
\left[1000,\dot{1}\dot{0}\dot{0}\dot{0}\right] &
\left[1001,\dot{1}\dot{0}\dot{0}\dot{1}\right] &
\left[1010,\dot{1}\dot{0}\dot{1}\dot{0}\right] &
\left[1011,\dot{1}\dot{0}\dot{1}\dot{1}\right] \\
\left[1100,\dot{1}\dot{1}\dot{0}\dot{0}\right] &
\left[1101,\dot{1}\dot{1}\dot{0}\dot{1}\right] &
\left[1110,\dot{1}\dot{1}\dot{1}\dot{0}\right] &
\left[1111,\dot{1}\dot{1}\dot{1}\dot{1}\right]
\end{pmatrix}}
\]

Let us consider now an explicit construction of the Maxwell field
$(1,0)\cup(0,1)$ within the quotient algebra
${}^\epsilon\C_4\cup{}^\epsilon\overset{\ast}{\C}_4$. First of all,
let us define a spinor representation of the field
$(1,0)\cup(0,1)$. As a rule, a spinor field of the quotient algebra
${}^\epsilon\C_4\simeq\cl_{4,1}$ is appearred at the extraction of the
minimal left ideal \cite{FRO90a,FRO90b,RSVL}:
\begin{equation}\label{left}
I_{4,1}=\cl_{4,1}e_{41}\simeq\cl^+_{1,3}e_{13}\frac{1}{2}(1+i\gamma_1\gamma_2)=
\ar\begin{pmatrix}
\psi_1 & 0 & 0 & 0\\
\psi_2 & 0 & 0 & 0\\
\psi_3 & 0 & 0 & 0\\
\psi_4 & 0 & 0 & 0
\end{pmatrix},
\end{equation}
where $e_{13}\frac{1}{2}(1+\gamma_0)$ and $e_{41}=\frac{1}{2}(1+\gamma_0)
\frac{1}{2}(1+i\gamma_1\gamma_2)$ are primitive idempotents of the
algebras $\cl_{1,3}$ and $\C_4$. Further, since $\cl^+_{1,3}\simeq\cl_{3,0}
\simeq\C_2$, then the spinor field $\psi$ in (\ref{left}) can be expressed
via the Dirac--Hestenes spinor field $\phi\in\cl_{3,0}\simeq\C_2$. Let
\[
\nabla=\partial^{0}\e_{0}+\partial^{1}\e_{1}+\partial^{2}\e_{2}+
\partial^{3}\e_{3},\quad
A=A^{0}\e_{0}+A^{1}\e_{1}+A^{2}\e_{2}+A^{3}\e_{3}
\]
be linear elements of the algebra $\cl_{3,0}$, where $A_i$ are the components
of the electromagnetic four--potential. Then
\begin{multline}\label{e10}
\nabla A=(\partial^{0}\e_{0}+\partial^{1}\e_{1}+
\partial^{2}\e_{2}+\partial^{3}\e_{3})(A^{0}\e_{0}+A^{1}\e_{1}+A^{2}\e_{2}+
A^{3}\e_{3})=\\
(\underbrace{\partial^{0}A^{0}+\partial^{1}A^{1}+\partial^{2}A^{2}+
\partial^{3}A^{3}}_{E^{0}})\e_{0}+(\underbrace{\partial^{0}A^{1}+
\partial^{1}A^{0}}_{E^{1}})\e_{0}\e_{1}+\\
(\underbrace{\partial^{0}A^{2}+\partial^{2}A^{0}}_{E^{2}})\e_{0}\e_{2}+
(\underbrace{\partial^{0}A^{3}+\partial^{3}A^{0}}_{E^{3}})\e_{0}\e_{3}+
(\underbrace{\partial^{2}A^{3}-\partial^{3}A^{2}}_{H^{1}})\e_{2}\e_{3}+\\
(\underbrace{\partial^{3}A^{1}-\partial^{1}A^{3}}_{H^{2}})\e_{3}\e_{1}+
(\underbrace{\partial^{1}A^{2}-\partial^{2}A^{1}}_{H^{3}})\e_{1}\e_{2}.
\end{multline}
The scalar part $E_{0}\equiv 0$, since the first bracket in (\ref{e10}) is
a Lorentz
condition $\partial^{0}A^{0}+\mbox{div}{\bf A}=0$. It is easy to see that 
other brackets are components of electric and magnetic fields:
$-E^{i}=-(\partial^{i}A^{0}+\partial^{0}A^{i})$, $H^{i}=
(\mbox{curl\bf A})^{i}$.

Since $\omega=\e_{123}$ belongs to a center of $\cl_{3,0}$, then
$$\omega \e_{1}=\e_{1}\omega=\e_{2}\e_{3},\quad\omega \e_{2}=\e_{2}
\omega=\e_{3}\e_{1},
\quad\omega \e_{3}=\e_{3}\omega=\e_{1}\e_{2}.$$

In accordance with these relations we can write (\ref{e10}) as follows
\begin{equation}\label{e11}
\nabla A=(E^{1}+\omega H^{1})\e_{1}+(E^{2}+\omega H^{2})
\e_{2}+(E^{3}+\omega H^{3})\e_{3}
\end{equation}
Further, let us compose the product 
$\nabla\bF$, where
$\bF$ is an 
expression
of the type (\ref{e10}):
\begin{multline}
\nabla\bF=\mbox{div\bf E}\e_{0}-((\mbox{curl\bf H})^{1}-
\partial^{0}E_{1})\e_{1}-((\mbox{curl\bf H})^{2}-\partial^{0}E^{2})\e_{2}-\\
-((\mbox{curl\bf H})^{3}-\partial^{0}E^{3})\e_{3}+((\mbox{curl\bf E})^{1}+
\partial^{0}H^{1})\e_{2}\e_{3}+((\mbox{curl\bf E})^{2}+\partial^{0}H^{2})
\e_{3}\e_{1}+\\
((\mbox{curl\bf E})^{3}+\partial^{0}H^{3})\e_{1}\e_{2}+\mbox{div\bf H}
\e_{1}\e_{2}\e_{3}.
\end{multline}
It is easy to see that the first coefficient of the product 
$\nabla\bF$
is a left part of the equation $\mbox{div\bf E}=\varrho$. The following three
coefficients compose a left part of the equation $\mbox{curl\bf H}-
\partial^{0}\mbox{\bf E}=j$,
other coefficients compose the equations 
$\mbox{curl\bf E}+\partial^{0}\mbox{\bf H}=0$
and $\mbox{div\bf H}=0$, respectively.

Further, since the element $\gamma_5=\gamma_0\gamma_1\gamma_2\gamma_3$
commutes with all other elements of the biquaternion (\ref{173}) and
$\gamma^2_5=-1$, then we can rewrite (\ref{173}) as follows
\[
\phi=(a^0+\gamma_{0123}a^{0123})+(a^{01}+\gamma_{0123}a^{23})\gamma_{01}+
(a^{02}+\gamma_{0123}a^{31})\gamma_{02}+(a^{03}+\gamma_{0123}a^{12})\gamma_{03}
\]
or taking into account (\ref{e11}) we obtain
\[
\phi=(E^1+iH^1)\gamma_{01}+(E^2+iH^2)\gamma_{02}+(E^3+iH^3)\gamma_{03}=
F_1\gamma_{01}+F_2\gamma_{02}+F_3\gamma_{03}.
\]
In the matrix form we have
\[
\phi=\ar\begin{pmatrix}
0 & 0 & F_3 & F_1-iF_2\\
0 & 0 & F_1+iF_2 & -F_3\\
F_3 & F_1-iF_2 & 0 & 0\\
F_1+iF_2 & -F_3 & 0 & 0
\end{pmatrix}
\]
Then from (\ref{left}) the relation immediately follows between spinors
$\psi\in\C_4$ and $\phi\in\cl_{3,0}$
\begin{equation}\label{S1}
\psi=\phi\frac{1}{2}(1+\gamma_0)\frac{1}{2}(1+i\gamma_1\gamma_2)=
\ar\begin{pmatrix}
0 & 0 & 0 & 0\\
0 & 0 & 0 & 0\\
F_3 & 0 & 0 & 0\\
F_1+iF_2 & 0 & 0 & 0
\end{pmatrix}\sim
\ar\begin{pmatrix}
0\\
F_1\\
F_2\\
F_3
\end{pmatrix}
\end{equation}
Let us consider now an action of the antiautomorphism 
$\cA\rightarrow\widetilde{\cA}$ on the arbitrary element of $\cl_{3,0}$
represented by a formula
\begin{equation}\label{Arb}
\cA=(a^0+\omega a^{123})\e_0+(a^1+\omega a^{23})\e_1+(a^2+\omega a^{31})\e_2
+(a^3+\omega a^{12})\e_3.
\end{equation}
The action of the antiautomorphism $\cA\rightarrow\widetilde{\cA}$ on the
homogeneous element $\cA$ of a degree $k$ is defined by a formula
$\widetilde{\cA}=(-1)^{\frac{k(k-1)}{2}}\cA$. Thus, for the element
(\ref{Arb}) we obtain
\[
\cA\longrightarrow\widetilde{\cA}=(a^0-\omega a^{123})\e_0+
(a^1-\omega a^{23})\e_1+(a^2-\omega a^{31})\e_2+(a^3-\omega a^{12})\e_3.
\]
Therefore, under action of $\cA\rightarrow\widetilde{\cA}$
the element (\ref{e11}) takes a form
\[
\widetilde{(\nabla A)}=(E^1-iH^1)\e_1+(E^2-iH^2)\e_2+(E^3-iH^3)\e_3
\]
and
\[
\widetilde{\phi}=\ar\begin{pmatrix}
0 & 0 & \overset{\ast}{F}_3 & \overset{\ast}{F}_1-i\overset{\ast}{F}_2\\
0 & 0 & \overset{\ast}{F}_1+i\overset{\ast}{F}_2& -\overset{\ast}{F}_3\\
\overset{\ast}{F}_3 & \overset{\ast}{F}_1-i\overset{\ast}{F}_2 & 0 & 0\\
\overset{\ast}{F}_1+i\overset{\ast}{F}_2 & -\overset{\ast}{F}_3 & 0 & 0
\end{pmatrix}.
\]
Hence it immediately follows
\begin{equation}\label{S2}
\widetilde{\psi}=\widetilde{\phi}\frac{1}{2}(1+\gamma_0)\frac{1}{2}
(1+i\gamma_1\gamma_2)=\ar\begin{pmatrix}
0 & 0 & 0 & 0\\
0 & 0 & 0 & 0\\
\overset{\ast}{F}_3 & 0 & 0 & 0\\
\overset{\ast}{F}_1+i\overset{\ast}{F}_2 & 0 & 0 & 0
\end{pmatrix}\sim
\begin{pmatrix}
0\\
\overset{\ast}{F}_1\\
\overset{\ast}{F}_2\\
\overset{\ast}{F}_3
\end{pmatrix}
\end{equation}
From (\ref{S1}) and (\ref{S2}) it follows that the full representation
space $\dS_4\cup\dot{\dS}_4$ is reduced to a 3--dimensional symmetric
space $\Sym_{(2,0)}\cup\Sym_{(0,2)}$. The transition from operator spinors
to $SO(3)$ vectors has been done by several authors (see, for example
\cite{Par00}).
The `vectors' (spintensors) of the
spaces $\Sym_{(2,0)}$ and $\Sym_{(0,2)}$ are
\begin{gather}
f^{00}=\xi^0\otimes\xi^0,\quad f^{01}=f^{10}=\xi^0\otimes\xi^1=
\xi^1\otimes\xi^0,\quad f^{11}=\xi^1\otimes\xi^1,\nonumber\\
f^{\dot{0}\dot{0}}=\xi^{\dot{0}}\otimes\xi^{\dot{0}},\quad
f^{\dot{0}\dot{1}}=f^{\dot{1}\dot{0}}=\xi^{\dot{0}}\otimes\xi^{\dot{1}}=
\xi^{\dot{1}}\otimes\xi^{\dot{0}},\quad
f^{\dot{1}\dot{1}}=\xi^{\dot{1}}\otimes\xi^{\dot{1}}.\label{Sp}
\end{gather}
It is well--known \cite{Rum36,RF68,RF} that spintensors (\ref{Sp})
correspond to a Helmholtz--Silberstein representation $\bF=\bE+i\bH$ and
form a basis of the Majorana--Oppenheimer quantum electrodynamics
\cite{Maj,Opp31,MRB74,Gia85,Rec90,Esp98,Dvo97} in which the electromagnetism
has to be considered as the wave mechanics of the photon.
Moreover, the field $(1,0)\cup(0,1)$ satisfies the Weinberg Theorem.

In such a way, it is easy to see that
\begin{eqnarray}
\text{Weyl--Hestenes neutrino field}&&(1/2,0)\cup(0,1/2),\nonumber\\
\text{Dirac electron--positron field}&&(1/2,0)\oplus(0,1/2),\nonumber\\
\text{Maxwell electromagnetic field}&&(1,0)\cup(0,1)\sim\nonumber\\
&&(1/2,0)\otimes(1/2,0)\cup(0,1/2)\otimes(0,1/2)\label{Max}
\end{eqnarray}
have the same mathematical structure. Namely, all these fields are obtained
from the fundamental field $(1/2,0)$ ($(0,1/2)$) by means of unification,
direct sum and tensor product. From this point of view a division of all
the physical fields into `gauge fields' and `matter fields' has an
artificial character. It is obvious that an origin of such a division
takes its beginning from the invalid description 
(from group theoretical viewpoint)
of electromagnetic field
in quantum field theory (a la Gupta--Bleuler approach) and a groundless
extension of the Yang--Mills idea (as known, a long derivative is equivalent
to a minimal interaction in the lagrangian). Further, as follows from
(\ref{Max}) the Maxwell field has a composite structure that gives rise
to a de Broglie--Jordan neutrino theory of light \cite{Bro32,Jor35},
in which the electromagnetic field is constructed via the two neutrino fields.

As known, the neutrino theory of light has a long and dramatic history
(see excellent review \cite{Dvo99,Per00}). The main obstacle that decelerates
the development of the neutrino theory of light is a Pryce's Theorem
\cite{Pry38} (see also a Berezinskii modification of this Theorem \cite{Ber66}).
In 1938, Pryce claimed the incompatibility of obtaining
transversely--polarized photons and Bose statistics from neutrinos.
However, as it has been recently shown by Perkins \cite{Per00} the
Pryce's Theorem contains an assumption which is unsupported and probably
incorrect. The Perkins arguments based on the experimental similarities
between bosons and quasi--bosons allow to overcome this obstacle
(Pryce's Theorem). In the present paper we show that transversely--polarized
photons follow naturally from composite neutrinos (it is a direct consequence
of the group theoretical consideration). It is obvious that such photons
are quasi--bosons \cite{Per01} that do not satisfy the Bose statistics,
but they satisfy the Lipkin statistics of a composite particle such as
the deuteron and Cooper pairs \cite{Lip73}.
In conclusion, one can say that a correct description of electromagnetic
field is a synthesis of the Majorana--Oppenheimer quantum electrodynamics
and de Broglie--Jordan neutrino theory of light.
\section*{Acknowledgments} I am grateful to Prof. E. Recami for
sending me unpublished Majorana works.


\begin{thebibliography}{00}
{\small
\bibitem{BS93} N.N. Bogoliubov, D.V. Shirkov, {\rm Quantum Fields} 
(Moskow, Nauka, 1993).
\bibitem{Bro32} L. de Broglie, Compt. Rend. {\bf 195}, 862 (1932). 
\bibitem{Ber66} V.S. Berezinskii, Zh. Ehksp. Theor. Fiz. {\bf 51}, 1374 (1966).
\bibitem{CF97} J.S.R. Chisholm, R.S. Farwell, {\it Properties of
Clifford Algebras for Fundamental Particles}, in {\rm Clifford (Geometric)
Algebras}, ed. W. Baylis (Birkhauser, 1996), pp. 365--388.
\bibitem{Dvo97} V.V. Dvoeglazov, {\it The Weinberg Formalism and a New Look
at the Electromagnetic Theory}, in The Enigmatic Photon. Vol.~IV (Eds.:
M. Evans, J.--P. Vigier, S. Roy and G. Hunter) p.~305--353, Series
Fundamental Theories of Physics. Vol.~90 (Ed. A. van der Merwe) Dordrecht,
Kluwer Academic Publishers 1997.
\bibitem{Dvo99} V.V. Dvoeglazov, {\it Speculations on the Neutrino Theory
of Light}, Annales de la Fondation de Louis de Brogle {\bf 24}, No. 1-4,
pp.111--128 (1999).
\bibitem{Esp98} S. Esposito, {\it Covariant Majorana Formulation of
Electrodynamics}, Found. Phys. {\bf 28}, 231--244 (1998).
\bibitem{FRO90a} V.L. Figueiredo, W.A. Rodrigues, Jr., E.C. Oliveira,
{\it Covariant, algebraic, and operator spinors}, Int. J. Theor. Phys. {\bf 29},
371-395, (1990).
\bibitem{FRO90b} V.L. Figueiredo, W.A. Rodrigues, Jr., E.C. Oliveira,
{\it Clifford algebras and the hidden geometrical nature of spinors},
Algebras, Groups and Geometries {\bf 7}, 153-198, (1990).
\bibitem{GMS} I.M. Gel'fand, R.A. Minlos, Z.Ya. Shapiro, Representations
of the Rotation and Lorentz Groups and their Applications (Pergamon Press,
Oxford, 1963).
\bibitem{Gia85} E. Giannetto, {\it A Majorana-Oppenheimer Formulation of
Quantum Electrodynamics}, Lettere al Nuovo Cimento {\bf 44}, 140--144 (1985).
\bibitem{Hest66} D. Hestenes, {\rm Space--Time Algebra} (Gordon \&
Breach, New York, 1966).
\bibitem{Hest67} D. Hestenes, {\it Real spinor fields}, J. Math.
Phys. {\bf 8}, 798-808, (1967).
\bibitem{Jor35} P. Jordan, Z. Phys. {\bf 93}, 434 (1935).
\bibitem{Lip73} H.J. Lipkin, Quantum Mechanics (North-Holland, Amsterdam,
1973) Chap. 6.
\bibitem{Lou93} P. Lounesto, {\it Clifford algebras and Hestenes
spinors}, Found. Phys. {\bf 23}, 1203-1237 (1993). 
\bibitem{Maj} E. Majorana, {\it Scientific Papers}, unpublished, deposited at
the ``Domus Galileana'', Pisa, quaderno {\bf 2}, p.101/1; {\bf 3}, p.11, 160;
{\bf 15}, p.16;{\bf 17}, p.83, 159.
\bibitem{MRB74} R. Mignani, E. Recami, M. Baldo, {\it About a Dirac-Like
Equation for the Photon according to Ettore Majorana}, Lettere al Nuovo
Cimento {\bf 11}, 568--572 (1974).
\bibitem{Nai58} M.A. Naimark, Linear Representations of the Lorentz Group
(Pergamon, London, 1964).
\bibitem{Opp31} J.R. Oppenheimer, Phys. Rev. {\bf 38}, 725 (1931).
\bibitem{Par00} J.M. Parra, {\it In what sense is the Dirac-Hestenes equation
a representation of the Poincar\'{e} group?}, in `Lorentz Group, CPT and
Neutrinos', A.E. Chubykalo et al Eds. World Scientific, Singapoore, 2000, p.370.
\bibitem{Per00} W.A. Perkins, {\it Interpreted History Of Neutrino Theory
Of Light And Its Future}, in `Lorentz Group, CPT and Neutrinos', 
A.E. Chubykalo et al Eds. World Scientific, Singapoore, 2000, p.115.
\bibitem{Per01} W.A. Perkins, {\it Quasibosons}, preprint hep-th/0107003 (2001).
\bibitem{Pry38} M.H.L. Pryce, Proc. Roy. Soc. (London) {\bf A165}, 247 (1938).
\bibitem{Rec90} E. Recami, {\it Possible Physical Meaning of the Photon
Wave-Function, According to Ettore Majorana}, in Hadronic Mechanics and
Non-Potential Interactions (Nova Sc. Pub., New York, 1990), pp. 231--238.
\bibitem{RSVL} W.A. Rodrigues, Jr., Q.A.G. de Souza, J. Vaz, Jr.,
P. Lounesto, {\it Dirac-Hestenes spinor fields in Riemann-Cartan spacetime},
Int. J. Theor. Phys., {\bf 35}, 1849-1900, (1996).
\bibitem{Rum36} Yu.B. Rumer, Spinorial Analysis (Moscow, 1936) [in Russian].
\bibitem{RF68} Yu.B. Rumer, A.I. Fet, {\it Polarization operator of quantized
fields}, Zh. Ehksp. Theor. Fiz. {\bf 55}, 1390--1392 (1968).
\bibitem{RF} Yu.B. Rumer, A.I. Fet, Group Theory and Quantized Fields
(Moscow, 1977) [in Russian].
\bibitem{Var99} V.V. Varlamov, {\it Fundamental Automorphisms of
Clifford Algebras and an Extension of D\c{a}browski Pin Groups},
Hadronic J. {\bf 22}, 497--535 (1999).
\bibitem{Var00} V.V. Varlamov, {\it Discrete Symmetries and Clifford Algebras},
Int. J. Theor. Phys. {\bf 40}(4), 769--805 (2001).
\bibitem{Var03} V.V. Varlamov, {\it Clifford Algebras and Discrete
Transformations of Spacetime}, Proceedings of Third Siberian Conference on
Mathematical Problems of Spacetime Physics, Novosibirsk, 20--22 June 2000.
(Institute of Mathematics Publ., Novosibirsk, pp.~97--135, 2001). 
\bibitem{Var01a} V.V. Varlamov, {\it Discrete Symmetries on the Quotient
Representation Spaces of the Lorentz Group}, Mathematical Structures and
Modelling {\bf 7}, 114--127 (2001).
\bibitem{Var01} V.V. Varlamov, {\it Clifford Algebras and Lorentz Group},
preprint math-ph/0108022 (2001).
\bibitem{Wein} S. Weinberg, {\it Feinman rules for any spin I {\rm\&} II
{\rm\&} III}, Phys. Rev. {\bf 133B}, 1318--1332 \& {\bf 134B}, 882--896 (1964)
\& {\bf 181B}, 1893--1899 (1969).
}
\end{thebibliography}
\end{document}